\documentclass{iaus}
\usepackage{graphicx}

\def\kms{~km~s$^{-1}$}
\def\etal{{\it et al.}}

\def\arcmin{$^{\prime}$}
\def\arcsec{$^{\prime\prime}$}
\def\msun{$M_\odot$}

\def\aj{\textit{AJ}~}
\def\apj{\textit{ApJ}~}
\def\apjl{\textit{ApJ} (Letters)~}
\def\aap{\textit{A\&A}~}

\def\mnras{\textit{MNRAS}~}

\def\nat{\textit{Nature}~}

\title[ALFALFA HI Clouds in Virgo] %% give here short title %%
{HI Clouds detected towards Virgo with the Arecibo Legacy Fast ALFA Survey}

\author[B. R. Kent]   %% give here short author list %%
{Brian R. Kent$^1$%
}

\affiliation{$^1$Center for Radiophysics and Space Research, Space Science Building
Cornell University, Ithaca, NY 14853, USA\break email: bkent@astro.cornell.edu\\[\affilskip]
}

\pubyear{2007}
\volume{244}  %% insert here IAU Symposium No.
\pagerange{1--10}
\date{?? and in revised form ??}
\jname{Dark Galaxies and Lost Baryons}
\editors{J. I. Davies, M. D. Disney, eds.}
\begin{document}

\maketitle

\begin{abstract}
The Arecibo Legacy Fast ALFA survey is in the process of yielding a complete HI dataset 
of the Virgo Cluster and its environs (Giovanelli et al. 2007, Kent et al., in preparation).
Assuming a distance to Virgo of 16.7 Mpc, the minimum detectable HI mass by ALFALFA is of order 
$2 \times 10^7$ M$_\odot$.  A number of the HI detections appear to have interesting properties.
Some appear associated with, but offset from, low surface brightness optical counterparts;
others, at larger spatial offsets, may be tidally related to optical counterparts.
Yet another class includes detections which are not identifiable with any optical counterparts.  
We present the ALFALFA results on these objects in the Virgo region, 
as well as followup aperture synthesis observations obtained with the VLA.
\end{abstract}

\firstsection % if your document starts with a section,
\section{Introduction}

Regions of high galaxian density
in the local Universe afford the opportunity
to explore a wealth of morphological diversity and physical processes.
Observations of the local Universe provide a census of
the low luminosity and low mass populations and their kinematic
and photometric properties.  Covering large areas of sky with 
blind surveys is extremely useful in giving a homogeneous dataset
that can be compared and correlated with observations at multiple wavelengths.

The Virgo Cluster is the nearest rich cluster of galaxies, at a distance
of 16.7 Mpc.  The $\sim$1300 cataloged
member galaxies of Virgo are morphologically segregated, with
early type galaxies inhabiting the cluster centers,
dominated by  ellipticals M87 and M49, and late type spirals
dominating the outer periphery.  Dwarfs are the most abundant class in Virgo, 
with over 850 members.  Binggeli, Sandage \& Tammann (1985) carried out
the first high quality optical survey of the cluster, releasing a catalog (VCC; Binggeli, Sandage, \& Tammann 1987, 1993)
that has served as an important standard of
reference for Virgo studies.
Redshift measurements show
that several subclumps in the outer periphery lie behind the main cluster centered around M87.
As the cluster progresses toward a state of virial equilibrium,
many galaxies in the outlying periphery of Virgo are moving at high speed while
falling into the cluster (Binggeli, Sandage \& Tammann~ 1993; Solanes \etal~ 2002).  The cluster
environment affects galaxies, evidenced by HI deficiency studies (Haynes \&
Giovanelli 1984; Solanes \etal~ 2002), in the form of ram pressure interaction
with the hot intracluster gas and other gravitational and hydrodynamic mechanisms.  High resolution
HI imaging has revealed disturbances and asymmetries in the gas disks, further
evidence that the galaxies are interacting with the cluster environment (Cayatte 1990; Chung 2007).

The 21cm spectral line of neutral hydrogen is an extremely useful
tool in determining the properties of galaxies.  Redshifts, kinematics,
and masses of gas content can all be obtained from global HI spectra.
As such, large area blind HI surveys can provide invaluable statistics
and information 
on the HI content of the local Universe, probing the faint end 
of the HI mass function within a cosmologically
fair volume, revealing information about high mass
galaxies, and tracing large scale structure.  The Arecibo 
Legacy Fast ALFA (ALFALFA)
extragalactic HI survey
is specifically designed to provide a large homogeneous dataset
of HI measurements for $cz_{\odot}\lesssim$ 18000 \kms~, able to detect as low as
$M_{HI}\sim 2 \times 10^7$ \msun~ at the distance of Virgo.
The blind survey will cover a total of 7000 deg$^2$ of high galactic 
latitude sky visible from Arecibo, including the full
region of the Virgo cluster and its environs, as well as many other cluster and and groups 
in the local Universe.  ALFALFA will serve as an important
second generation HI survey to the successful HIPASS survey (Barnes \etal~ 2001),
improving in sensitivity, bandwidth, and spectral resolution.

\section{ALFALFA Observations of the Virgo Cluster region}

The Virgo cluster and its surrounding periphery are a primary target
for ALFALFA, as the cluster is easily accessible to the Arecibo telescope.  
An effort has been made to prioritize and complete coverage of that region by ALFALFA.
The current Virgo catalogs encompass data taken in the Spring 2005
and 2006 observing sessions (Giovanelli \etal~2007 and Kent \etal~ in preparation).
ALFALFA is conducted with the 7-element ALFA receiver in meridian 
transit drift mode.  Spectra are sampled at a rate of 1 Hz.  
A 100 MHz bandwidth with 4096 channels yields a spectral
resolution of 24.4 kHz (5.1\kms~ at the 21cm HI line).  Drift scans
are recorded in ten minute intervals, with a calibration noise
diode signal inserted into the system between scans .
After conversion into the IDL environment,
software routines developed at Cornell are used for automated bandpass calibration
and user-controlled radio frequency interference flagging
and quality monitoring.  Completed drifts are then combined
into regularly gridded data cubes that encompass 2.4 $\times$ 2.4 degrees
of sky.  The data cubes are split into 1024 channel 
redshift sections overlapping by $\sim$1000\kms~ in redshift space.
Signal extraction is completed in Fourier space using
an automated matched-filter algorithm (Saintonge 2007)
and candidate detections are confirmed by eye in 
a custom made visualization environment.  Detections
are corroborated with optical imaging from the Digital
Sky Survey and SDSS; fluxes and $cz_{\odot}$ measurements
are extracted from integrated spectral profiles.  Published
survey catalogs and spectra 
are available at http://arecibo.tc.cornell.edu/hiarchive/alfalfa/.
In addition, National Virtual Observatory protocols allow catalog access via other
websites through a cone search service.  The whole data processing
pipeline of ALFALFA is embedded in an environment that facilitates cross-referencing
into other data sources.

\begin{figure}[t]
\centering
\includegraphics{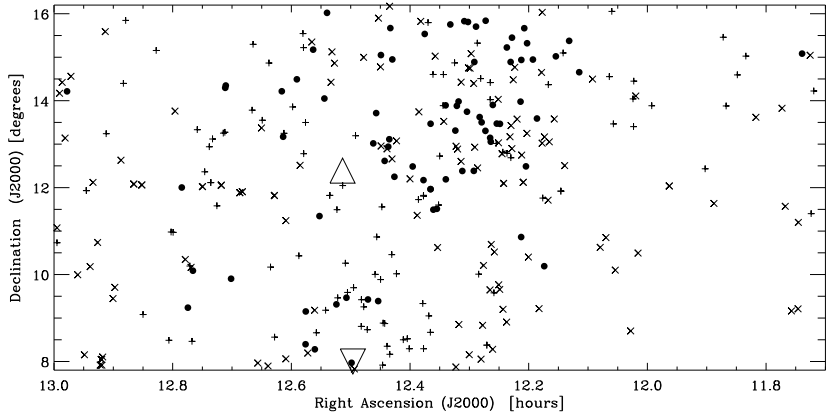}
  \caption{Sky distribution of galaxies for $cz_{\odot} < 3000$ \kms in the Virgo region so far
           detected by the ALFALFA survey.  Large cluster galaxies M87 and M49 are indicated by the upward 
           and downward pointing open triangles respectively.  Filled circles indicate galaxies
           with $cz_{\odot} <$500\kms, plus signs(+) indicate galaxies with 500$< cz_{\odot} <$1500\kms, and $\times$ symbols
           indicate galaxies with $cz_{\odot} >$1500\kms.}
\label{fig:skyplot}
\end{figure}

The data presented in this paper encompass $\sim 240 $~deg$^2$ of sky
from  11$^h$44$^m <$ R.A.(J2000) $< 14^h 00^m$ and 08$^{\circ}$00\arcmin $<$ Dec.(J2000) $< 16^{\circ} 00$\arcmin.
Over 1300 sources were detected out to $cz_{\odot} \sim 18,000$\kms.  The sky distribution
of galaxies for $cz_{\odot} < 3000$\kms around the Virgo core region is shown in Figure 2. 
In this region, ALFALFA detects 5.5 sources deg$^{-2}$.
A velocity distribution of the catalog
is depicted in Figure 2.  The dark histogram in the same
figure shows the galaxies detected for the same
area of sky by the HIPASS Survey (Wong \etal~ 2006).
The overdensity of sources near $cz_{\odot} \sim 1000$\kms~
indicates the Virgo cluster population.  The histogram clearly
shows large scale structure beyond Virgo, most prominently with the 
Coma supercluster regime at $cz_{\odot}\sim$ 7000\kms.  
Statistics of this dataset are shown in Figure 3, displaying
the signal-to-noise and flux integral versus velocity width of the HI sources.  
The signal-to-noise limit of 6.5 is shown in the bottom plot as a function
of velocity width; it is expected to increase as the square root of the velocity width (dotted 
line)
for widths $\lesssim$~300 \kms and linearly for larger widths, as observed.

When compared with the remainder of the survey, the Virgo
area component of ALFALFA will allow for the comparison between the HI mass function
in the cluster and in the field.  However, because of large peculiar velocities
towards the Virgo region, primary distance indicators (TRGB, SBF) will be needed
to determine an accurate cluster HI mass function.
Figure 4 shows the detections' HI mass vs distance, determined by using the
results of mass flow models (Tonry \etal~ 2000; Masters 2005), 
except for objects within the Virgo cluster region, which are all placed at a distance of 16.7 Mpc.

%Several comparisons can be made at this point between HIPASS and ALFALFA.  
%64 HIPASS sources are detected in this region for 
%the HIPASS bandpass coverage(Wong \etal 2006); ALFALFA's 
%sensitivity, bandpass coverage, and spectral resolution 
%have yielded the detection of 1332 sources. 

\begin{figure}
\centering
 \includegraphics{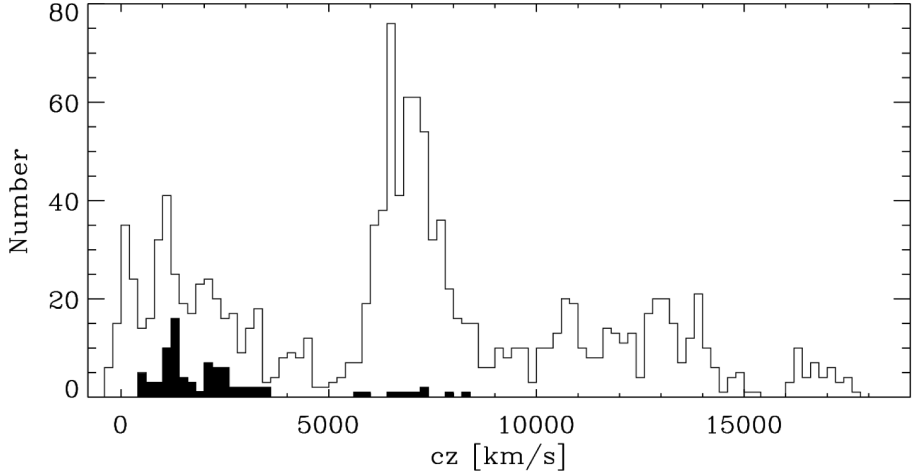}
  \caption{$cz_{\odot}$ distribution of detections for the region outlined in Figure 1.  The dark histogram
           shows the distribution of HIPASS sources for the same region (Wong \etal~ 2006).
           Notable features showing large scale structure include the Virgo cluster population
           at $cz_{\odot}\sim 1000$\kms~ and the Coma supercluster regime at $cz_{\odot}\sim$7000\kms.}
           \label{fig:czhist}
\end{figure}

\begin{figure}[t]
\centering
 \includegraphics{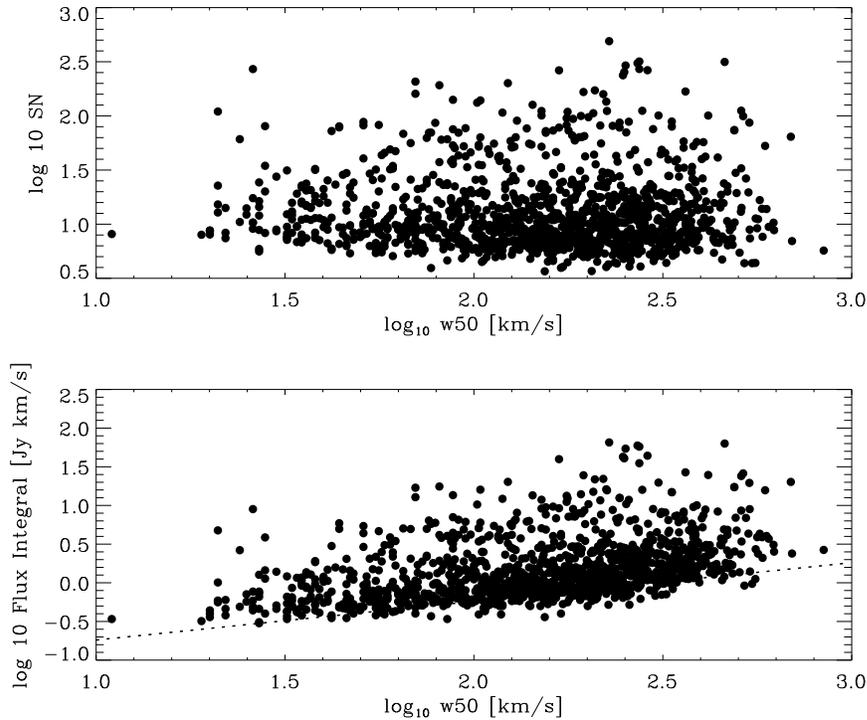}
  \caption{The top figure shows the S/N vs. velocity width for the Virgo region sources
           detected with ALFALFA. 
           The bottom figure shows the integrated flux vs. velocity width, with the dotted
           line indicating a S/N limit of 6.5.}\label{fig:scatterstat}
\end{figure}

\begin{figure}[t]
\centering
 \includegraphics{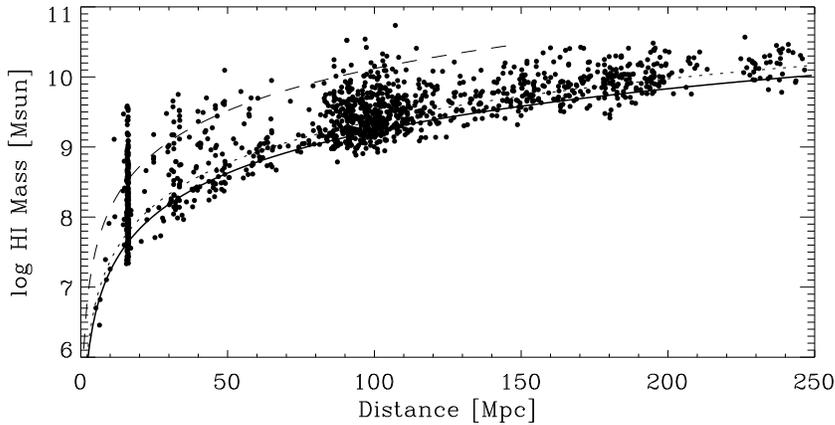}
  \caption{HI mass vs. distance for the Virgo region sources presented in this paper.
           Assumed Virgo cluster members are seen as a vertical line near 16.7 Mpc.
           The top dashed line indicates a flux integral of 5.6 Jy\kms, corresponding
           to a HIPASS 5$\sigma$ detection.
           The middle dotted line indicates a flux integral of 1.0 Jy\kms,
           and the lowest solid line indicates 0.72 Jy\kms limit for the ALFALFA survey.
           All three lines are computed with a source velocity width $W_{\rm{FWHM}}=200$\kms.}
           \label{fig:himass}
\end{figure}

\section{ALFALFA HI Clouds in Virgo}

A small
subgroup of objects have been found to have
no apparent optical counterparts
in the first Virgo portion of the ALFALFA survey (Kent \etal~ 2007).
Here we report on eight detections, several with multiple components.  With
the exception of one, all are within the canonical sky and
$cz_{\odot}$ bounds of the
Virgo Cluster periphery defined in Binggeli, Sandage \& Tammann (1993).  
Other objects of interest are discussed in the presentations
of Giovanelli and Koopmann in these proceedings.
The HI measurements
and data of these clouds are listed in Table 1.  The Table 1 columns
are described as follows:\\

\noindent\textit{Col.(1)} - Cloud ID number\\
\textit{Col.(2 \& 3)} - HI source center coordinates (J2000); these positions are typically
accurate to within 24\arcsec or better (see Giovanelli \etal~ 2007)\\
\textit{Col.(4)} - Heliocentric velocity in \kms\\
\textit{Col.(5)} - Velocity width measured at half peak power in \kms\\
\textit{Col.(6)} - Integrated flux in Jy \kms\\
\textit{Col.(7)} - Signal to noise ratio\\
\textit{Col.(8)} - Base 10 logarithm of the HI mass in solar units, assuming HI is optically thin\\
\textit{Col.(9)} - Angular distance from M87 in degrees\\  %Column 10 removed by BK

Figure 5 shows these HI detections overlaid 
on a smoothed ROSAT X-ray map (Snowden \etal~ 1995).
These optically ``dark'' objects are not detected in large numbers towards
Virgo, and no detections sharing similar characteristics have
been found towards the ALFALFA anti-Virgo region (Saintonge \etal~ 2007).
In the high galaxian density environment of Virgo and its periphery, ram pressure 
stripping and high speed interaction are likely to take place, and gas removal from disks
is more frequent.

Galaxy harassment also has been put forth
as a strong influence on cluster evolution; galaxies can be stripped
of their gas during high speed encounters with other cluster
member and the cluster's potential (Moore \etal~ 1996).  Many
large disk galaxies inhabiting the outer parts of Virgo show evidence
of having undergone some sort of dynamical interaction.  HI studies of
selected galaxies have shown that the HI disks are disturbed and/or
asymmetric.

Previous studies have shown examples
where the fingerprint of cluster interaction
can clearly be seen.  An HI plume extending from NGC~4388 ($cz_{\odot}$ = 2524 \kms; 
Davies \etal~ 2004; Oosterloo \& van Gorkom 2005)
was also detected in the ALFALFA survey (Table 1, Cloud 4).  This feature 
has been attributed
to interaction between the galaxy and the hot intracluster gas.
Another example is listed as Cloud 5, likely a tidal interaction
between dwarf irregular UGC~7636 and M49, the large galaxy central
to the Virgo B subcluster.

Another object of note is Cloud 6.  Originally
detected with the Jodrell Bank telescope(Davies \etal~ 2004; Minchin \etal~ 2005a)
and dubbed VIRGOHI21,
this detection has been tidally linked to the nearby large spiral NGC~4254.  Galaxy
harassment likely plays a role in the formation of such a tail, which extends
over 250 kpc north of the associated spiral (Haynes, Giovanelli \& Kent 2007).
A more detailed description can be found in these proceedings by Giovanelli.
Also presented are the results of modeling of the interaction that my have caused it
in Duc (these proceedings) and Duc \& Bournaud (2007).

Cloud 8 is discussed here although the associated SB0/a 
galaxy NGC~4795 ($cz_{\odot}$ = 2781 \kms) and dwarf NGC~479 6($cz_{\odot}$ = 2406 \kms) lie
behind Virgo.  
The presence of anomalous HI in this system was noted by Hoffman \etal~ (1989) and
Duprie \& Schneider (1996).  The ALFALFA map shows the HI is
in clouds lying outside the optical disks of these galaxies.

The remaining objects in Table 1 (Clouds 1,2,3, \& 7) are isolated, in that
they are not obviously connected with large galaxies in their respective vicinities,
nor can they be seen to be associated with faint but visible optical counterparts.
Cloud group 7, detected during
the Spring 2005 observing runs.  The complex is composed of five components, stretched
over 200 kpc at the Virgo distance, situated between M87 and M49 on the sky.
Moment zero maps of the complex showing the components are shown in Figure 6.
Each map is integrated over 50 km/s, centered at the velocities shown
in the upper right corners.
Maps show flux contours at the 120, 200, 300, 400, 500, 600, 700, 800, 900, and 1000 mJy beam$^{-1}$~\kms
~flux levels.  The velocity dispersion of the system is $\sim$250 \kms~ in extent.
Aperture synthesis observations of this system have been obtained with the VLA and are
in preparation for publication (Kent \etal~ 2007).

Clouds 1, 2, \& 3 share similar characteristics
to each other, in that they are narrow sources ($W_{\rm{FWHM}} \lesssim 50$\kms~),
unresolved by the Arecibo beam, 
and all located in the western periphery of Virgo.  
At the Virgo distance, these objects are of order 10 kpc
or less in diameter.  This region is also
home to a population making up the so-called ``M'' Cloud (Binggeli, Sandage, \& Tammann 1993).
This subgroup is thought to lie behind the main A cluster around M87;
the velocity of the ``M'' Cloud population with respect to the systemic
Virgo velocity also suggests that many of these galaxies are currently
falling into the main cluster.  These objects are far removed
from the cluster core, and it is unlikely that hydrodynamical
processes are the main instigator of this detachment from spiral disks.  Instead, gravitational interaction
with the cluster itself or high speed encounters are more likely to produce such debris.
Such action within the cluster periphery can yield higher probabilities of tidal
debris being found in this region.  However, it remains a possibility
that these objects could be associated with very small lower surface brightness
dwarf galaxies or be veritable ``dark galaxies''.  Further optical followup will shed light on the issue.

\begin{figure}[t]
\centering
 \includegraphics[width=5.0in]{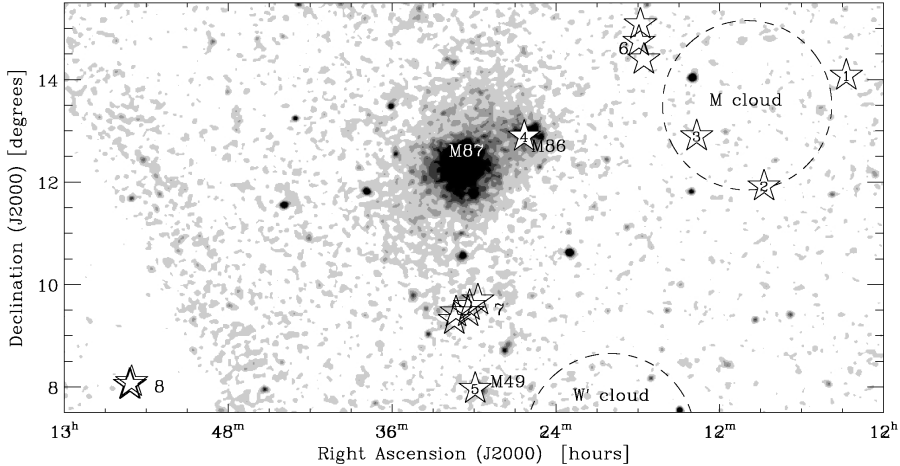}
  \caption{Optically unseen HI detections presented in this paper.  The 
           ROSAT X-ray background has been smoothed with a 5\arcmin~ Gaussian
           kernel.  The M and W\arcmin clouds are outlined by dashed circles
           (Binggeli, Popescu, \& Tammann (1993).  The numbering corresponds
           to the cloud identifiers given in Column 1 of Table 1.}\label{fig:xray}
\end{figure}

Analysis of aperture synthesis observations obtained 
with the VLA has shown that Cloud 2 can be resolved.
The experiment utilized 12 hours of observation in C-configuration,
centered on the Arecibo centroid for Cloud 2 given in Table 1.
%This centroid position was observed for twelve 30 minute intervals
%with 5 minute phase calibrator observations in between each
%on source observation.  In addition, 0542+498 was used
%as a flux calibrator, observed at the beginning of the run.
The backend setup consisted of a 1.56 MHz bandwidth centered at 
$cz_{\odot}=1230$ \kms~, with 128 channels, yielding a spectral resolution
of 12.2 kHz (2.5 \kms~at the 21cm HI line).  
%Standard AIPS reduction
%was utilized; the dataset was flagged for bad baselines and any
%stray RFI.  The data were bandpass calibrated, and the dirty 
%synthesized beam pattern was deconvolved from the dataset
%using the CLEAN algorithm (H{\"o}gbom 1974; Schwarz 1978).  
A data cube
using uniform weighting was created, with pixel elements smoothed
to 10\arcsec.  A preliminary moment map was made from this cube,
and is shown in Figure 7.  The map rms is $\sigma=$3.8 mJy beam$^{-1}$~\kms,
with contours indicating 3, 4, 5, 6, 7, 8 and 9$\sigma$.  The 
large cross indicates the centroid of the Arecibo 
detection (3.8\arcmin $\times$ 3.3\arcmin~ beam).  The source
is resolved into three extended clumps.  The two main clumps are
of high S/N, located near the pointing
center and appear to be connected by a bridge.  The peaks
are separated by $\sim$2\arcmin~ in space ($\sim$9 kpc at the Virgo
distance).  A third detection lies $\sim$3\arcmin~ southeast
of the northern most clump.  The small, compact group
of HI detections composing Cloud 2 
are spread over no more than 15\arcmin~ to 20\arcmin~ in extent,
with a velocity spread of $\sim$20 \kms.
The aperture synthesis data reveals that Cloud 2 
is similar to the Cloud 7 complex, albeit on different size scales.
Aperture synthesis observations have also been obtained for other
cloud complexes in Table 1 and will be presented in
a later publication (Kent \etal~2007).

Simple timing arguments can be employed to explore the dynamics
of both these systems.  The velocity differences between the components
of Cloud 7 are $\sim$250\kms~.  Assuming that the global complex
is not gravitationally bound, then the rate of separation between
the cloud elements would be $\sim$250 kpc Gyr$^{-1}$.  A similar scenario
is found with Cloud 2.  A 20\kms~ separation between the narrow, compact
elements yields a separation rate of $\sim$20 kpc Gyr$^{-1}$.  In both cases,
the clouds will disperse in a gigayear or less, well within a cluster crossing time.
This is indicative that these objects could be transient in nature, which
could indicate why they are not detected in large numbers.

\section{Summary}

An overview of the core Virgo cluster catalog and a listing of detections
with no apparent optical counterparts have been presented.  The catalog
shows a vast improvement on the first generation of blind HI surveys
with hundreds of new HI detections, greater sensitivity, and both
spectral and spatial resolution.  HI clouds that cannot
be matched to optical counterparts are not detected in
large numbers.  Those that are detected
lie in the Virgo periphery.
ALFALFA and VLA maps reveal kinematics about some of the 
multi-component detections, suggesting that they can easily disperse
within a cluster crossing time.

\begin{acknowledgments}

The Arecibo Observatory is part of the National Astronomy
and Ionosphere Center which is operated by Cornell University
under a cooperative agreement with the National Science Foundation.

The National Radio Astronomy Observatory is facility of the National
Science Foundation operated under cooperative agreement by 
Associated Universities, Inc.

This research has made use of the NASA/IPAC Extragalactic Database (NED) which is 
operated by the Jet Propulsion Laboratory, California Institute of Technology, 
under contract with the National Aeronautics and Space Administration.  
This work has been supported by NSF 
grants AST--0307661, AST--0435697, and AST--0607007.

Funding fo the SDSS and SDSS-II has been provided by the Alfred P. Sloan
Foundation, the Participating Institutions, the National Science Foundation,
the U.S. Department of Energy, the National Aeronautics and Space 
Administration, the Japanese Monbukagakusho, the Max Planck Society,
and the Higher Education Council for England.  The SDSS website
is http://www.sdss.org/.

{\it Skyview} was developed and 
maintained under NASA ADP Grant NAS5--32068 under the auspices of the High 
Energy Astrophysics Science Archive Research Center at the Goddard Space 
Flight Center Laboratory of NASA.

This research has made use of data obtained from or software provided by the 
US National Virtual Observatory, which is sponsored by the National Science Foundation.

\end{acknowledgments}

\begin{figure}[t]
\centering
 \includegraphics{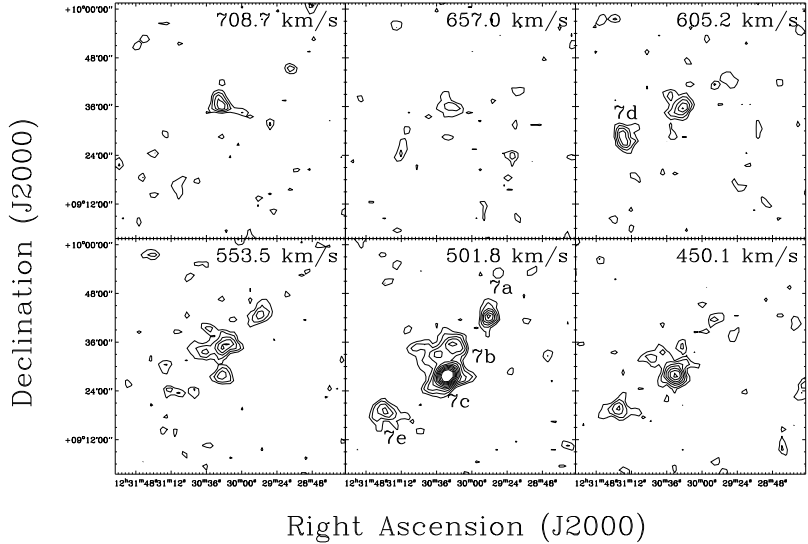}
  \caption{ALFALFA moment zero HI maps for Cloud 7 listed in Table 1.
           Each map is integrated over 50\kms~ centered at the $cz_{\odot}$ value 
           indicated in the upper right corners.  Map contours
           are for 120, 200, 300, 400, 500, 600, 700, 800, 900, and 1000 mJy beam$^{-1}$~\kms~
           flux levels.}\label{fig:VirgocomplexAO}
\end{figure}

\begin{figure}[t]
\centering
 \includegraphics{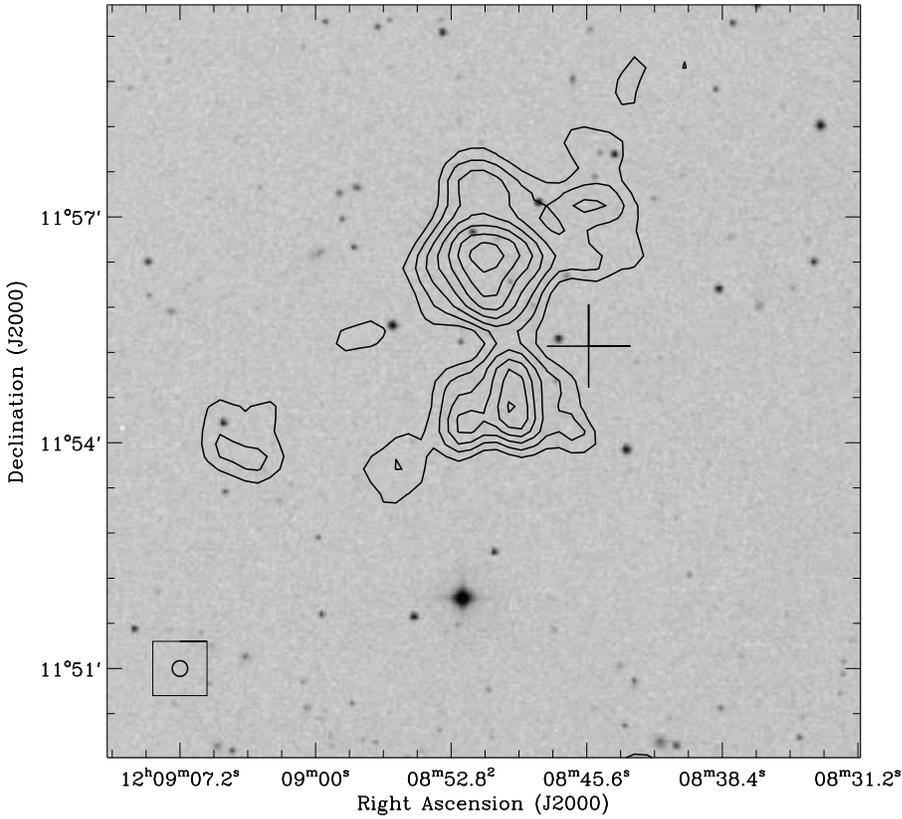}
  \caption{VLA C-array moment zero HI map of Cloud 2 from Table 1.  The map was
           created using uniform weighting.  The map rms is $\sigma=$3.8 mJy beam$^{-1}$~ \kms,
           with contours indicating 3, 4, 5, 6, 7, 8 and 9$\sigma$.  The 
           large cross indicates the centroid of the Arecibo detection (3.8\arcmin $\times$ 3.3\arcmin beam).  
	   The synthesized beam size is indicated in the lower left corner.
           }\label{fig:vlamapCloud2}
\end{figure}

%Example Table

%\begin{table}[bht]
%  \caption{A sample table illustrating usage of the \LaTeX{} table
%    environment. Please try, if possible, to use the smaller font size
%    for the table, as in the current example (set by the {\tt
%      footnotesize} command).}
%  \label{MSU_tab:table}
%  \begin{center}
%    \leavevmode
%    \footnotesize
%    \begin{tabular}[h]{lrcc}
%      \hline \\[-5pt]
%      First column & Col. 2      &  Col. 3      &  V mag \\[+5pt]
%      \hline \\[-5pt]
%      row 1  & 11.0 & 25.0 & 12 \\
%      row 2  & 11.0 & 25.0 & 12 \\
%      row 3  & 11.0 & 25.0 & 12 \\
%      row 4  & 11.0 & 25.0 & 12 \\
%      row 5  & 11.0 & 25.0 & 12 \\
%      \hline \\
%      \end{tabular}
%  \end{center}
%\end{table}

\begin{table}
\caption{ALFALFA:  Optically Unseen detections towards Virgo}
  \begin{center}
    \leavevmode
    \footnotesize
    \begin{tabular}[h]{rrrrccrcl}
      \hline \\[-5pt]

              Cloud ID
	    & $\alpha$
	    & $\delta$ 
	    & $cz_\odot$ 
	    & $W_{\rm{FWHM}}$
	    & $F_{c}$
	    & S/N 
	    & log$_{10} M_{HI}$
	    & d$_{M87}$ \\[+5pt]
	    %Units
	     
	      Units
	    & J2000 
	    & J2000 
	    & (\kms) 
	    & (\kms) 
	    & (Jy \kms)  
	    & ---
	    & \msun
	    & deg\\

      \hline \\[-5pt]
       1$^{\rm{a}}$ & 12 02 44.4 & +14 04 56  & 1121$\pm$  1  &   22$\pm$  2  &   0.30 $\pm$ 0.02 &   5.1   & 7.29 & 7.0\\
       2$^{\rm{b}}$ & 12 08 45.5 & +11 55 17  & 1230$\pm$  1  &   29$\pm$  2  &   0.77 $\pm$ 0.04  & 11.6   & 7.63 & 5.4\\
       3$^{\rm{b}}$ & 12 13 41.8 & +12 53 51  & 2235$\pm$  2  &   53$\pm$  3  &   1.21 $\pm$ 0.07  &  9.2   & 8.54 & 4.2 \\
       4$^{\rm{c}}$ & 12 26 19.4 & +12 53 30  & 2246$\pm$  5  &  135$\pm$ 11  &   2.05 $\pm$ 0.07  & 14.4   & 8.77 & 1.2 \\
       5$^{\rm{d}}$ & 12 29 54.7 & +07 58 12  &  473$\pm$  5  &   30$\pm$ 10  &   1.13 $\pm$ 0.06  & 10.9   & 7.87 & 4.4 \\
       6{\it a}$^{\rm{e}}$ & 12 17 55.5 & +14 44 45  & 1984$\pm$  1  &  128$\pm$  2  &   2.09 $\pm$ 0.06  & 16.2   & 8.70 & 3.9 \\
       6{\it b}$^{\rm{e}}$ & 12 17 49.1 & +15 04 52  & 2200$\pm$  6  &   40$\pm$ 13  &   0.52 $\pm$ 0.05  &  5.0   & 8.16 & 4.1 \\
       6{\it c}$^{\rm{e}}$ & 12 17 33.8 & +14 23 47  & 2111$\pm$  10 &   65$\pm$ 20  &   0.57 $\pm$ 0.04  &  7.3   & 8.17 & 3.8 \\
       7{\it a}$^{\rm{b}}$ & 12 29 42.8 & +09 41 54  &  524$\pm$  7  &  116$\pm$ 15  &   1.16 $\pm$ 0.07  &  8.6   & 7.87 & 2.7   \\
       7{\it b}$^{\rm{b}}$ & 12 30 19.4 & +09 35 18  &  603$\pm$  4  &  252$\pm$  7  &   2.56 $\pm$ 0.09  & 13.1   & 8.22 & 2.8 \\
       7{\it c}$^{\rm{b}}$ & 12 30 25.8 & +09 28 01  &  488$\pm$  5  &   62$\pm$ 11  &   2.48 $\pm$ 0.07  & 21.2   & 8.21 & 2.9 \\
       7{\it d}$^{\rm{b}}$ & 12 31 19.0 & +09 27 49  &  607$\pm$  4  &   56$\pm$  7  &   0.72 $\pm$ 0.06  &  6.5   & 7.67 & 2.9 \\
       7{\it e}$^{\rm{b}}$ & 12 31 26.7 & +09 18 52  &  480$\pm$ 10  &   53$\pm$ 21  &   0.91 $\pm$ 0.06  &  7.6   & 7.77 & 3.1 \\
       8{\it a}$^{\rm{f}}$ & 12 55 04.3 & +08 06 13  & 2629$\pm$  3  &   71$\pm$  7  &   0.72 $\pm$ 0.07  &  6.4   & 8.43 & 7.3 \\
       8{\it b}$^{\rm{f}}$ & 12 55 10.2 & +08 02 44  & 2754$\pm$ 14  &  407$\pm$ 27  &   2.91 $\pm$ 0.12  &  9.4   & 9.08 & 7.4 \\
       8{\it c}$^{\rm{f}}$ & 12 55 13.7 & +08 02 51  & 2771$\pm$  4  &  292$\pm$  7  &   2.52 $\pm$ 0.10  & 10.9   & 9.02 & 7.4 \\
    
      \hline \\

      \end{tabular}

      \begin{tabular}[h]{l}

      a) Confirmed by follow-up, high sensitivity observation at Arecibo.\\ 
      b) VLA maps obtained, processing underway.\\
      c) VirgoHI4 (Davies \etal 2004), VLA map by Oosterloo \etal (2005).\\
      d) Vicinity of M49, Sancisi \etal (1987); synthesis data by Henning \etal (1993).\\
      e) Clumps in VirgoHI21, WSRT data by Minchin \etal (2005).\\
      f) NGC 4795/4796 group.\\

      \hline \\
      \end{tabular}
      
  \end{center}
\end{table}

%AASTeX version
%\tablenotetext{a}{Confirmed by follow-up, high sensitivity observation at Arecibo.}
%\tablenotetext{b}{VLA maps obtained, processing underway.}
%\tablenotetext{c}{VirgoHI4 (Davies \etal 2004), VLA map by Oosterloo \etal (2005).}
%\tablenotetext{d}{Vicinity of M49, Sancisi \etal (1987); synthesis data by Henning \etal (1993).}
%\tablenotetext{e}{Clumps in VirgoHI21, WSRT data by Minchin \etal (2005).}
%\tablenotetext{f}{NGC 4795/4796 group.}

\end{document}